\documentclass[sigconf,authorversion]{acmart}

\usepackage[normalem]{ulem}
\usepackage{multirow}
\usepackage{adjustbox}
\usepackage{array}
\usepackage{booktabs} 
\usepackage{soul}
\usepackage{color}
\usepackage{colortbl}
\usepackage{cleveref}
\usepackage{siunitx}
\usepackage{subfigure}
\usepackage{graphicx}
\usepackage{caption}

\definecolor{lightcyan}{HTML}{80cdc1}
\definecolor{lightbrown}{HTML}{dfc27d}

\newcommand{\visual}{\cellcolor{lightcyan}}
\newcommand{\textual}{\cellcolor{lightbrown}}

\usepackage{fontawesome5}
\usepackage{tikzsymbols}
\usepackage{twemojis}
\usepackage{halloweenmath}
\usepackage{enumitem}
\newcolumntype{R}[2]{%
    >{\adjustbox{angle=#1,lap=\width-(#2)}\bgroup}%
    l%
    <{\egroup}%
}

\AtBeginDocument{%
  \providecommand\BibTeX{{%
    \normalfont B\kern-0.5em{\scshape i\kern-0.25em b}\kern-0.8em\TeX}}}

\copyrightyear{2024} 
\acmYear{2024} 
\setcopyright{acmcopyright} 

\copyrightyear{2024} 
\acmYear{2024} 
\setcopyright{acmlicensed}\acmConference[SIGIR '24]{Proceedings of the 47th International ACM SIGIR Conference on Research and Development in Information Retrieval}{July 14--18, 2024}{Washington, DC, USA}
\acmBooktitle{Proceedings of the 47th International ACM SIGIR Conference on Research and Development in Information Retrieval (SIGIR '24), July 14--18, 2024, Washington, DC, USA}
\acmDOI{10.1145/3626772.3657768}
\acmISBN{979-8-4007-0431-4/24/07}

\begin{document}

\title[Explainability for Transparent Conversational Information-Seeking]{Explainability for Transparent Conversational Information-Seeking}

\author{Weronika Łajewska}
\orcid{https://orcid.org/0000-0003-2765-2394}
\affiliation{%
  \institution{University of Stavanger}
  \city{Stavanger}
  \country{Norway}
}
\email{weronika.lajewska@uis.no}

\author{Damiano Spina} 
\orcid{https://orcid.org/0000-0001-9913-433X}
\affiliation{
\institution{RMIT University}
  \city{Melbourne}
  \country{Australia}
}
\email{damiano.spina@rmit.edu.au}

\author{Johanne Trippas} 
\orcid{https://orcid.org/0000-0002-7801-0239}
\affiliation{
\institution{RMIT University}
  \city{Melbourne}
  \country{Australia}
}
\email{j.trippas@rmit.edu.au}

\author{Krisztian Balog}
\orcid{https://orcid.org/0000-0003-2762-721X}
\affiliation{%
  \institution{University of Stavanger}
  \city{Stavanger}
  \country{Norway}
}
\email{krisztian.balog@uis.no}

\begin{abstract}
The increasing reliance on digital information necessitates advancements in conversational search systems, particularly in terms of information transparency. While prior research in conversational information-seeking has concentrated on improving retrieval techniques, the challenge remains in generating responses useful from a user perspective. This study explores different methods of explaining the responses, hypothesizing that transparency about the source of the information, system confidence, and limitations can enhance users' ability to objectively assess the response. By exploring transparency across explanation type, quality, and presentation mode, this research aims to bridge the gap between system-generated responses and responses verifiable by the user. We design a user study to answer questions concerning the impact of (1) the quality of explanations enhancing the response on its usefulness and (2) ways of presenting explanations to users. The analysis of the collected data reveals lower user ratings for noisy explanations, although these scores seem insensitive to the quality of the response. Inconclusive results on the explanations presentation format suggest that it may not be a critical factor in this setting. 

\end{abstract}



\maketitle

\section{Introduction}
The field of conversational information-seeking (CIS) focuses on systems designed for dialogue-based information retrieval, where the aim is to enable interactions that closely resemble human conversation~\citep{Zamani:2023:FNT}. 
In recent years, research in this space has primarily concentrated on improving various components of the response generation process, such as passage retrieval, reranking,  query rewriting, and on making answers self-contained~\citep{Zamani:2023:FNT, Owoicho:2022:TRECb, Sekulic:2024:CHIIR}. However, it remains a challenge to create a trustworthy conversational response from the retrieved information~\citep{Ren:2021:TOIS}. 
In transitioning from traditional search engine result pages to conversational information-seeking systems that limit responses to a few sentences, there is a significant concealment of underlying details such as the ranking of results and specifics about the sources. These details are essential for users to assess the scope, novelty, reliability, and topical relevance of the provided information~\citep{Xu:2006:J.}.
Recently, retrieval-augmented generation (RAG) has been proposed~\citep{Lewis:2020:NIPS}, which is claimed to produce more factually correct and diverse content. RAG, however, does not solve issues around transparency, as it is not able to indicate low-confidence responses or identify potential flaws related to limitations of the retrieved results or of the response generation process itself.
Since the user is provided only with a short textual response as the final outcome of the generation process, it becomes the responsibility of the conversational system to identify and communicate any potential limitations to its users, ensuring transparency and empowering users to evaluate response quality. While the importance of explainability is broadly recognized for AI~\citep{Monroe:2018:CommACM} and has been extensively studied, for example, for decision support and recommender systems~\citep{Nunes:2017:UMUAI,Zhang:2020:FnTIR}, it has not received due attention for CIS systems.

In this study, we aim to fill this gap by investigating approaches to explaining conversational responses, as a means to increase the transparency of the system.
Our focus is on \emph{informational transparency}, disclosing information about the limitations or potential pitfalls in the response generation needed to enable appropriate understanding and assessment, in contrast to \emph{functional} understanding of what the system can do, by exposing its capabilities and limitations or \emph{mechanistic} understanding focused on how the system works~\citep{Liao:2023:arXiv}.
In particular, the focus of this study lies on the sources used for generating the response, the system's confidence in the provided information, and potential limitations or pitfalls of the response. 
In contrast to prior research on reporting system confidence~\citep{Cau:2023:IUI, Rechkemmer:2022:CHIa} or identifying particular limitations~\citep{Zhong:2020:ACL, Huang:2019:CoNLLb}, our emphasis is on effectively communicating this information to the user in a conversational setting; we thus assume the existence of components that estimate system confidence and perform the detection of limitations.

\begin{figure}
    \centering
    \includegraphics[width=0.48\textwidth]{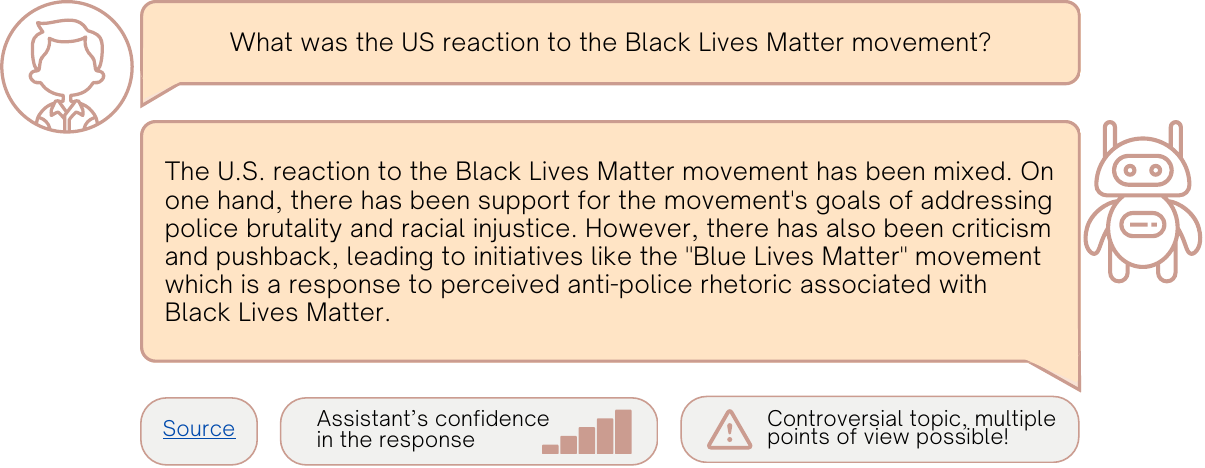}
    \captionsetup{aboveskip=5pt}
    \caption{Information-seeking dialogue with a CIS system with explanations (sources, confidence, and limitations).}
    \label{fig:system-user-dialogue}
    \vspace{-4mm}
\end{figure}

Specifically, based on the previous research in related domains, we choose to increase the transparency of a CIS system by explaining (1) the origin of presented information, i.e., source~\citep{Tsai:2021:CHI, Bohnet:2023:arXiv, Liu:2023:EMNLP}, (2) the system's confidence~\citep{Cau:2023:IUI, Radensky:2023:FAccT}, and (3) potential limitations of the generated response~\citep{Sakaeda:2022:NAACL-HLT}; see Figure~\ref{fig:system-user-dialogue} for an illustration of an enhanced conversational response. Being transparent about these aspects of the response can enable users to make informed judgments about the presented information and increase their perceived usefulness of the response. We investigate the user's perception of the system response quality together with the type and quality of explanations.
We ask the following two main research questions.

\emph{\textbf{RQ1: 
How does the quality of responses and explanations affect user-perceived response usefulness?}}
Individuals without specific training can only distinguish between human-generated and auto-generated texts at a level close to random chance~\citep{Clark:2021:ACL-IJNLP}. Indeed, users easily overlook factually incorrect, unsupported, biased, or incomplete information. 
Therefore, we investigate the impact of the quality of the response and explanations provided by the system on users' assessment of the response. Specifically, imperfect responses in our study include factual errors or lack of viewpoint diversification, while noisy explanations introduce problems related to sources (information subjectivity or lack of support for the response), confidence (incorrect scores), or limitations (irrelevant information about pitfalls).

\emph{\textbf{RQ2: What are effective ways to provide explanations to users?}}
There are multiple approaches to providing users with explanations. One option is incorporate them as part of the natural language system utterance, ensuring that users are explicitly informed about the confidence and potential pitfalls of the response
~\citep{Rechkemmer:2022:CHIa}. As an alternative, we explore utilizing various user interface elements to effectively convey the response's limitations~\citep{Lu:2021:CHIa, Shani:2013:Ja} or providing a granular scale of the system's confidence in generated response~\citep{Shani:2013:Ja}. Building on findings from studies in recommender systems and automated decision making~\citep{Nunes:2017:UMUAI,Zhang:2020:FnTIR}, we seek to adapt and explore these concepts within the context of CIS systems.

To answer the above questions, we conduct a crowdsourcing-based user study with 160 participants asking about their perception of responses and explanations that vary in quality and presentation mode.
The analysis of the collected data reveals that users can identify situations when their level of confidence in the response does not match the confidence reported by the system, but they cannot detect noise in the explanations of system limitations. In general, noisy explanations result in lower user-reported scores for response usefulness. In terms of presentation mode, we observe no significant advantage of any of the modes, which suggests that the explanation format is not critical for users.

Overall, our study seeks to establish a more trustworthy interaction in CIS dialogues by bridging the gap between system-generated responses and their usefulness to the users, by providing explanations.
The main contributions of this work include (1) the first user study that provides explanations of source, confidence, and limitations in the CIS domain, (2) a manually curated dataset of responses and explanations, with noise incorporated in a controlled manner, and (3) generalizable results describing the impact of noise and the presentation mode of the explanations on response usefulness and user ratings of explanations of source, confidence, and limitations. 

The paper is accompanied by an online repository, containing the manually generated CIS responses and explanations, user study results, and scripts for data analysis at \url{https://bit.ly/TransparentCIS}.

\section{Related Work}
\label{sec:related}

While traditional search engines provide ranked lists of documents, conversational response generation aims to aggregate top-ranked passages into coherent answers~\citep{Ren:2021:TOIS}. The task of generating summaries from retrieved results was first piloted in the 2022 edition of the TREC Conversational Assistance track (CAsT)~\citep{Owoicho:2022:TRECb}. Generative language models have been widely adopted for response generation~\citep{Zhang:2020:ACLa}; however, the final step of aggregating supporting facts through abstractive summarization has its challenges, including factual errors~\citep{Tang:2022:NAACL-HLT} and hallucinations~\citep{Ji:2023:ACMa, Tang:2023:ACL, Cao:2016:COLINGb}. 
Despite advancements, CIS systems still face limitations such as unanswerability~\citep{Sulem:2022:NAACL-HLTa, Choi:2018:EMNLPb, Rajpurkar:2018:ACLb, Reddy:2019:Trans.}, biases and lack of viewpoint diversification~\citep{Gao:2020:Information, Draws:2021:SIGIRa, Sakaeda:2022:NAACL-HLT, Azzopardi:2021:CHIIR}, or queries with impossible conditions~\citep{Huang:2019:CoNLLb, Lee:2020:LRECb}. Even though research has been done in related fields, such as text classification~\citep{Zhong:2020:ACL, Kim:2019:ECIR}, question answering~\citep{Liao:2022:SIGIR, Rajpurkar:2018:ACLb}, or reading comprehension~\citep{Zhang:2021:AAAI, Huang:2019:CoNLLb} 
towards detecting these issues, communicating detected problems to users is still a largely unexplored area in CIS. 
To ensure the transparency of the system, the response should disclose system capabilities and potential limitations, thereby managing user expectations~\citep{Radlinski:2017:CHIIRb, Azzopardi:2018:CAIR}. 

Unlike previous studies that concentrated on detecting limitations, our work emphasizes the effective communication of potential flaws in the system's output to the user. Such limitations can be revealed using natural language utterances~\citep{Rechkemmer:2022:CHIa}, using analogy~\citep{He:2023:Proc.}, incorporating user interface elements~\citep{Lu:2021:CHIa, Koch:2023:J}, or by providing a granular scale of the system's confidence~\citep{Shani:2013:Ja}. Studies in the context of recommender systems have demonstrated that reporting the system's confidence in predictions can offer valuable information to users, aiding them in making informed decisions~\citep{Shani:2013:Ja}. Moreover, research on human-AI interactions has highlighted the significance of system performance feedback in shaping human trust and reliance on AI systems~\citep{Lu:2021:CHIa, Rechkemmer:2022:CHIa}. Both the level of confidence displayed by machine learning models and their observed accuracy influence people's belief in the model's predictions, and users' willingness to follow the model's suggestions, especially after observing the model's performance in practice~\citep{Rechkemmer:2022:CHIa}. 

\begin{figure*}[t]
    \centering
    \includegraphics[width=\textwidth]{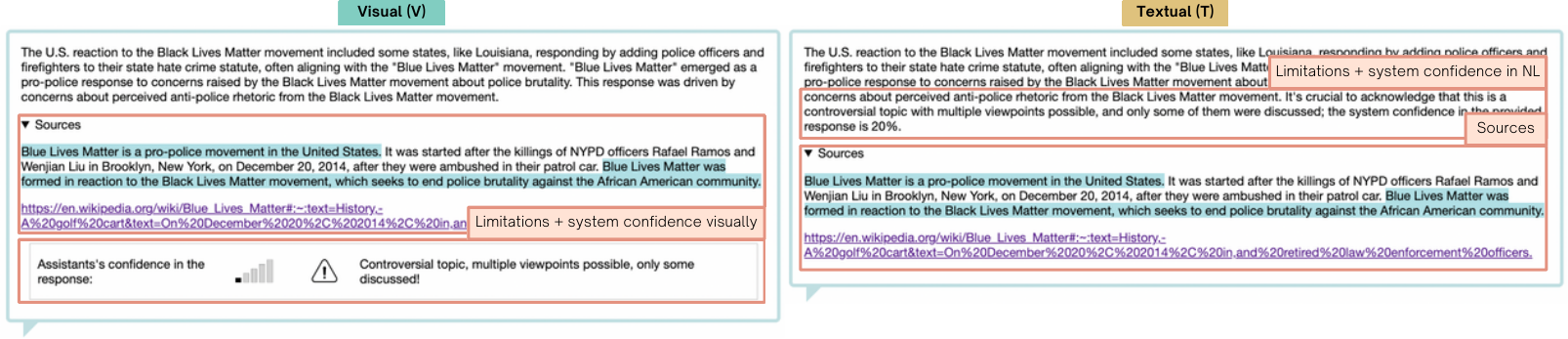}
    \captionsetup{aboveskip=-0pt}
    \caption{Examples of responses with explanations for the query: \emph{What was the US reaction to the Black Lives Matter movement?} 
    }
    \label{fig:presentation_modes}
    \vspace{-2mm}
\end{figure*}

Effective user-system interactions, aligning user expectations, and building trust in the systems that we attempt to achieve by communicating explanations about the response to the user are the main axes of explainable AI (XAI). According to human-AI interactions design guidelines, the system should communicate its capabilities, reliability, and the rationale behind its decisions~\citep{Amershi:2019:CHI}. XAI research in the context of decision-making emphasizes the role of explanations in improving user comprehension~\citep{Cheng:2019:CHI} and increasing human trust in the system~\citep{Zhang:2020:FAT}. Explanations can vary in terms of presentation format (textual vs. visual)~\citep{Zhang:2020:FAT}, the level of interactivity~\citep{Cheng:2019:CHI}, complexity~\citep{Tsai:2021:CHI}, or reasoning styles~\citep{Cau:2023:IUI}. Explanations can also be used to reveal the system's confidence~\citep{Cau:2023:IUI}, system accuracy indicators~\citep{Kocielnik:2019:CHI}, data sources~\citep{Tsai:2021:CHI}, answer attribution~\citep{Bohnet:2023:arXiv, Liu:2023:EMNLP} and the correctness of system suggestions~\citep{Cau:2023:IUI}. 
While explanations can enhance users' adherence to the system's advice, they may also lead to incorrect mental models of the systems---overconfidence or overreliance---especially among users with low domain expertise~\citep{Cau:2023:IUI}. Therefore, understanding the perceived usefulness of explanations is the first step in designing reliable and transparent CIS systems.

\section{Methodology}
\label{sec:methodology}

We aim to investigate the user's perception of the (1) system response quality, (2) type and quality of explanations, and (3) presentation of explanations.
We assume a \textit{retrieval-augmented generation system} that, given a query, performs the following steps: (1) it retrieves passages and identifies the information nuggets in the top retrieved results containing key pieces of information answering a user query; (2) it synthesizes the identified snippets (i.e., information nuggets) into a concise and natural language response; (3) it returns the system’s confidence in the provided response; and (4) based on the provided query, retrieved information nuggets, and returned confidence, it identifies the potential pitfalls and limitations that could have contributed to flaws in the response.
We consider three types of explanations the system may provide: 
(1) the underlying \emph{source}, to help users verify the response's factual correctness and broader context; (2) the system's \emph{confidence} in the provided response, to give users insights about how certain the outcome of response generation is; and (3) potential \emph{limitations or pitfalls} to warn the user about flaws in the response or the source.

The study's main goal is to investigate whether explanations provided by the system can make the user's response assessments easier or increase the information's usefulness. We provide crowd workers with different configurations of responses and explanations, varying in quality and presentation mode, and ask them to indicate their perception of different system response dimensions.
Inspired by work in the area of explainable decision-making systems~\citep{Cau:2023:IUI}, we explore two different ways of presenting the explanations about the response limitations and system confidence: textual and visual presentation; see~Figure~\ref{fig:presentation_modes}.

\subsection{Experimental Design}

We have defined ten experimental conditions using different variants of the response and explanations.\footnote{We acknowledge that the variants for each transparency dimension are not exhaustive. Various UI elements can be used to present information, and different ways to introduce noise can be explored. However, since response-related explanations have not been explored in conversational search, we limit the first study in this area to solutions previously proposed for similar systems.} 
Covering all combinations of factors (explanation components $\times$ quality $\times$ presentation mode) exhaustively would be unfeasible. Therefore, 
we select a subset of experimental conditions that best represent what we are trying to measure in our study. The selected conditions vary along three main dimensions: (1) response quality, (2) quality of the explanations (i.e., source, system confidence, limitations), and (3) presentation style (see~Table~\ref{tab:experimental_conditions}). More details about experimental conditions and the different explanation variants can be found in Section~\ref{sec:input_data}.

The ten experimental conditions resulted in ten different human intelligence tasks (HITs).
In each HIT, crowd workers are 
asked to assess responses for ten queries.
This is to ensure that the obtained results are to a large extent topic-independent.
To avoid repeated judgments that would reduce the reliability of the study, we allow each crowd worker to complete only one HIT~\citep{Steen:2021:ACL}, i.e., we employ a between-subject design~\citep{Kelly:2007:FNTa}.
In each HIT, the order of query-response pairs is intentionally randomized. This is done to prevent any adverse effects on the given query-response pairs that might occur if they were consistently presented towards the end of the task, where worker fatigue could potentially influence the results. 

\begin{figure}
    \centering
    \vspace*{-0.25\baselineskip}
    \includegraphics[width=0.35\textwidth]{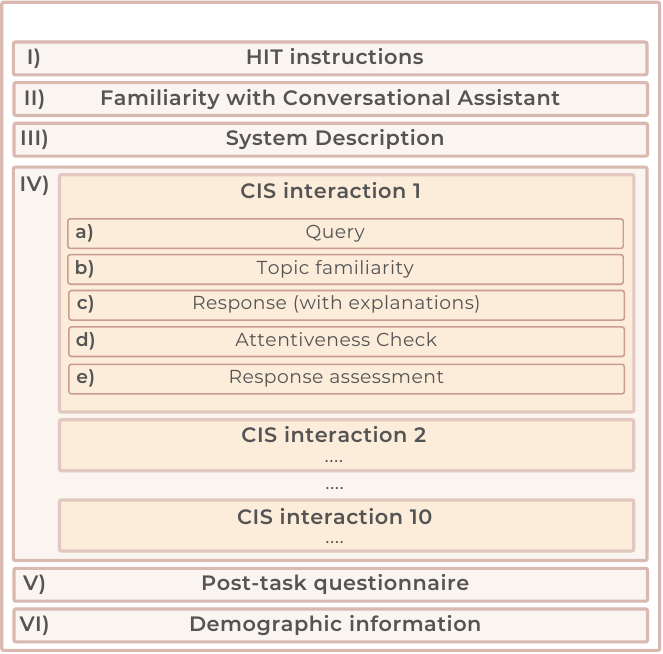}
    \captionsetup{aboveskip=2pt}
    \caption{High-level design of the user study.}
    \label{fig:task_components}
\end{figure} 

\subsection{Crowdsourcing Task Design}
Figure~\ref{fig:task_components} summarizes the design of the crowdsourcing tasks. Each HIT contains ten query-response pairs and is comprised of: I) HIT instructions providing task background; II) a questionnaire about the worker's familiarity with conversational assistants 
(see Table~\ref{tab:questionnaire-user_background}); III) a description of the system; IV) ten CIS interactions; V) a post-task questionnaire; and VI) a demographics questionnaire.  Workers are not given specific examples of query-response pairs in the instructions to avoid bias.
Part III contains a pre-use explanation of the system~\citep{Chiang:2022:IUI}. It aims at improving the following competencies of the users: (1) understanding the capabilities of the system, and (2) understanding that the response is limited to 3 sentences only. 
We decompose part IV of the user study into ten subsections using independent CIS interactions to facilitate atomic microtask crowdsourcing~\citep{Gadiraju:2015:IEEE}. Each CIS interaction contains (a) a query; (b) a topic familiarity questionnaire; (c) a system response possibly enhanced with explanations; (d) a corresponding attentiveness check; and (e) a CIS response assessment. 
CIS interactions are followed by a post-task questionnaire (Part V) investigating workers' experience of interacting with the assistant in general, not concerning specific responses. The questionnaire contains indirect questions about all three types of explanations enhancing the system response (see~Table~\ref{tab:questionnaire-user_background}).
The HIT finishes with a short demographics questionnaire (Part VI) asking workers' age, education level, and gender.\footnote{Specific questions posed in each user study can be found in the online repository.}

\subsubsection*{a) Query and b) Topic Familiarity}

The query is followed by a short questionnaire asking about interest, familiarity, and likelihood of posing a similar query~\citep{Bolotova:2020:CIKM} (see~Table~\ref{tab:questionnaire-user_background}). In this user study, the worker's background knowledge and familiarity with the topic are dependent variables that we cannot control. Asking users to assess their familiarity with the topic enables us to condition the collected data on users' background knowledge~\citep{Krishna:2021:NAACL-HLT}.

\subsubsection*{c) Response}

The system response synthesizes the information nuggets identified in the top retrieved results. The response can be enhanced with explanations that can be presented in different formats.

\subsubsection*{d) Attentiveness Check}

We present workers with an attentiveness check for each query-response pair, 
to detect poorly performing workers, cheat submissions, or bots~\citep{Gadiraju:2015:IEEE}. 
Each attention check consists of three sentences related to the topic of the query, one of them being a summary of the provided response. Sentences are provided in a random order and workers are asked to select the best summary~\citep{Bolotova-Baranova:2023:ACL}. This simple quality check enables us to filter out responses from workers who are not performing the task attentively or reading the responses carefully. Submissions that failed on more than 3 out of 10 attentiveness questions were rejected.

\subsubsection*{e) Response Assessment}
\label{subsec:response_assessment}

In this part of the CIS interaction, workers are asked to evaluate different dimensions of the response variant presented for a given query. The question about each response dimension is answered on a four-point Likert scale. Explicitly asking users to report on its value is not helpful because they may have a different understanding of this concept~\citep{Kelly:2007:FNTa}. Therefore, in our setup, user satisfaction is indirectly observable. To increase the ecological validity of our experiments, the questions do not use explicitly the names of the dimensions. Instead, we ask about each response dimension using an operational definition (see~Table~\ref{tab:response_dimensions}). This approach ensures a common understanding of the dimensions by all study participants. 
Both the response dimensions and the operational definitions are inspired by \citet{Cambazoglu:2021:CHIIRa}'s work investigating answer utility for non-factoid question answering. 

\begin{table}[tp]
    \captionsetup{skip=2pt}
    \caption{Operational definitions used in the response assessment questionnaire for all response dimensions. They followed a statement: \emph{The provided assistant's response \ldots} and were answered by crowd workers on a four-point Likert scale.}
    \label{tab:response_dimensions}
    \centering
    \small
    \adjustbox{max width=0.48\textwidth}{
    \begin{tabular}{p{2cm}p{7.6cm}}
    \toprule
         \textbf{Response \mbox{Dimension}} & \textbf{Operational definition used in the user study} \\
        \midrule
        Usefulness & \ldots was useful for completing my task \\
        \midrule
        Relevance & \ldots is about the subject of the question \\
        Correctness & \ldots contains an accurate response to the question  \\
        Completeness & \ldots covers every aspect of the question\\
        Comprehensiveness & \ldots contains detailed information \\
        Conciseness & \ldots does not contain redundant information \\
        Serendipity & \ldots contains some unexpected but positively surprising information \\
        Coherence &  \ldots does not contain inconsistent statement \\
        Factuality & \ldots is based on things that are known to be true \\
        Fairness & \ldots is free of any kind of bias  \\
        Readability & \ldots is fluently written \\
        Satisfaction & \ldots is satisfying in terms of completing my information need \\
        \bottomrule 
    \end{tabular}
    }
    \vspace{-4mm}
\end{table}

\begin{table}[tp]
    \captionsetup{skip=2pt}
    \caption{Questions used for collecting data about the user experience of using conversational agents, their involvement in the topic, and their rating for explanations.}
    \label{tab:questionnaire-user_background}
    \centering
    \small
    \adjustbox{max width=0.48\textwidth}{
    \begin{tabular}{p{3cm}p{7.4cm}}
    \toprule
         \textbf{Variable} & \textbf{Question used in the user study} \\
        \midrule
        Conversational Agent \mbox{Familiarity} & How often do you use conversational assistants like Siri, Alexa, or Google Assistant? \\
        Search with Agent Freq. & How often do you use conversational assistants to search for information? \\
        \midrule
        Topic Familiarity & What is your level of familiarity with the topic of the question? \\
        Interest in Topic & What is your level of interest in the question? \\
        Similar Search Probability & What is the likelihood that you would search for this information? \\
        \midrule
        Source \mbox{Explanation} & To what extent were the provided responses supported? \\
        Limitation \mbox{Explanation} & To what extent did the assistant help you realize the potential limitations of the responses? \\
        Confidence \mbox{Explanation} & To what extent are you aware of the assistant's confidence in the provided responses?\\
        \bottomrule 
    \end{tabular}
    }
    \vspace{-4mm}
\end{table}
\section{User Study Execution}
\label{sec:user_study_execution}

We used the Amazon Mechanical Turk (MTurk)
crowdsourcing platform to collect responses from online workers.\footnote{Our institution does not require ethics approval for this kind of studies.} Data collection was run between 20 December 2023 and 9 January 2024, divided into two stages: a pilot (Section~\ref{sec:pilot}) and a main study (Section~\ref{sec:on_scale}).

\subsection{Data}
\label{sec:input_data}

A critical element of the study is selecting query-response pairs and explanations enhancing the responses that enable us to answer our research questions. 
We use ten queries selected from the TREC CAsT 2020~\citep{Dalton:2020:TREC} and 2022~\citep{Owoicho:2022:TRECb} datasets and two manually created responses for each query. 
Different variants of the responses (perfect and imperfect) and explanations (accurate and noisy) are created manually by one of the authors of the paper. The noise in responses and explanations is introduced manually using framing, i.e., distorting the information presented to the users~\citep{Kocielnik:2019:CHI}. 
For each source, the specific information nuggets that contributed to the answer are highlighted, inspired by the CAsT-snippets dataset~\citep{Lajewska:2023:CIKM}.

\paragraph*{Queries and Responses}
\label{sec:queries_responses}

The query selection process takes into account the potential challenges of the query and the familiarity of crowd workers with the topic. We select a subset of queries from the TREC CAsT datasets that are challenging in one of two aspects: (1) limited coverage of the topic in the corpus or lack of a full answer, resulting in factual errors; or (2) topic complexity or controversy resulting in an incomplete or biased response. By selecting these challenging queries we attempt to simulate scenarios where enhancing the system response with explanation can be beneficial for users. Additionally, queries selected in the first step are sorted according to the familiarity scores reported by crowd workers in a small crowdsourcing study that was set up to select the top ten queries that are deemed most well-known to users. This approach aims to ensure that users possess sufficient background to meaningfully assess responses and associated explanations. 
We consider two variants of the response for each query: perfect and imperfect. The perfect response, i.e., ground truth answer, is generated manually using the top retrieved results by one of the authors of the paper. The imperfect response is a manual modification of the ground truth answer to contain factual errors, be biased towards one point of view, or cover only one aspect of a complex problem. This way, we attempt to take into account significantly different versions of the responses in terms of their accuracy and quality.

\paragraph*{Explanations}

We provide explanations related to (1) source, (2) system confidence, and (3)  limitations, which are instantiated in two variants: accurate and noisy.

\subsubsection*{(1) {Sources}}

The ``Source'' component is an expandable element within the response, encompassing the complete text of the paragraph used for generating the response. It includes annotations of information nuggets~\citep{Lajewska:2023:CIKM}, highlighting crucial pieces of information within the passage. Additionally, workers receive a link to the entire webpage from which the passage originates~\citep{Liu:2023:EMNLP}. This allows them to access the full text of the document, aiding in the assessment of its relevance, which is particularly beneficial for long, non-navigational queries~\citep{Kazai:2022:ECIR}. The URLs are anchored to the specific section of the webpage where the passage is located. Additionally, based on the URL, workers can assess the credibility or authority of the source.
The noisy source pertains to the query's topic but lacks information that supports the provided response~\citep{Liu:2023:EMNLP}. It corresponds to the initial passage from the Wikipedia page related to the general query topic, allowing for an assessment of users' diligence in verifying the provided explanations.

\subsubsection*{(2) {System Confidence}}

Within conversational response generation, confidence can be assessed along several different dimensions:
\begin{itemize}[leftmargin=3ex]
    \item The confidence that the identified snippets contain the full, complete answer to the question, not only part of it.
    \item Given that the response is limited to only 3 sentences, the confidence that the top-\textit{k} snippets used in the response provide a sufficient coverage of the retrieved information.
    \item The confidence that the response generated with LLM using the selected snippets is accurate; this accuracy is tied to the model's fluency in the topic, assessing how adept the model is in crafting content on a given subject, which can be influenced by the volume of data during LLM training or topic popularity.
\end{itemize}
System confidence is either communicated in textual form (\emph{``the system confidence in the provided response is ...\%''} appended at the end of the response)~\citep{Cau:2023:IUI} or through an additional UI element. Given that users best understand confidence displays inspired by well-known displays in other areas~\citep{Shani:2013:Ja}, we decided to use a bar chart presentation that is often associated with cell phone connectivity.
Adding noise to this component results in system confidence being reverted, i.e., although the provided response is correct, a low system confidence is reported. We consider the confidence of 1--2 out of 5 for imperfect responses and 4--5 for perfect responses. We skip confidence of 3 as it is ambiguous and we skip confidence of 0 as it represents the situation when the system should not show any response, but state that an answer could not been found.\footnote{Users' reactions to such extreme confidence scores is not a subject of this study, but could be explored in future work once it has been established that users find  confidence explanations useful.}
\subsubsection*{(3) {Response Limitations}}
We have identified several key areas of potential challenges and problems that could impact the usefulness of provided responses. These issues, while not exhaustive, serve as a starting point for consideration in our user study. Among the challenges related to the topic, we recognize the potential issues related to controversy, leading to a lack of viewpoint diversification, and complexity, resulting in response incompleteness. Source-related challenges include the subjectivity of the source text used, possibly outdated source information, sources influenced by commercial interest, promoting specific products, or brands, and reliance on unverified or not reputable sources. Query-related issues encompass biases or ambiguities in queries, time-sensitive queries requiring current information, queries lacking sufficient context, privacy-sensitive queries involving private or confidential information, and speculative queries seeking insights into future events. Additionally, search and system issues may arise, such as rare topics insufficiently covered in the corpus, lack of credible sources supporting the response, or retrieved passages containing contradictory information. 

The query-response pairs selected for this study contain factual errors, are incomplete, or rely on subjective sources. Additionally, challenges related to the topic, source, or query, not identified in the subset of query-passage pairs used in this study, are also considered to explore whether users can more easily identify these issues based on the presence and correctness of explanations provided by the system. Issues purely related to search or system failures, where the system is aware of its inability to find sources that answer the question, fall outside the scope of this study. In such cases, the system should inform the user about no answer found without trying to produce a response. Response limitations are communicated either in a textual form by appending running text at the end of the system response~\citep{Rechkemmer:2022:CHIa, Costa:2018:IUI} or using an additional UI element resembling a warning message (inspired by fact-checking warning labels~\citep{Koch:2023:J}).
Adding noise to limitation explanation results in communicating irrelevant limitations, i.e., if the topic is controversial, the system informs the user about query ambiguity or possibly outdated source information. We aim for the noisy limitations to be easily distinguishable after reading the query and the response carefully. The goal of this study is to investigate the limitation explanations rather than the detection of specific limitations, therefore the noise in the limitations is aimed to be easy to spot.

\begin{table}[tp]
    \centering
    \captionsetup{skip=2pt}
    \label{tab:components_variants}
      \caption{Experimental conditions considered in the user study; components may be included without noise (+); with some inaccuracies (\textasciitilde); or not provided in the system's output (--). \colorbox{lightbrown}{(T)} indicates textual and \colorbox{lightcyan}{(V)} visual presentation mode.}
    \adjustbox{max width=0.48\textwidth}{
    \begin{tabular}{lcccccccccccc}
    \toprule
           & \textbf{EC1} & \textbf{EC2} & \textbf{EC3} & \textbf{EC4} & \textbf{EC5} & \textbf{EC6} & \textbf{EC7} & \textbf{EC8} & \textbf{EC9} & \textbf{EC10} \\
        \midrule
         Response & \textual +, T & \textual +, T & \textual \textasciitilde, T & \textual \textasciitilde, T & \textual +, T & \textual +, T & \textual \textasciitilde, T & \textual \textasciitilde, T & \textual +, T & \textual \textasciitilde, T \\
         \midrule
         Source & \textual +, T & \textual +, T & \textual +, T & \textual +, T & \textual \textasciitilde, T & \textual \textasciitilde, T & \textual \textasciitilde, T & \textual \textasciitilde, T & -- & -- \\
         Confidence & \visual +, V & \textual +, T & \visual +, V & \textual +, T & \visual \textasciitilde, V & \textual \textasciitilde, T & \visual \textasciitilde, V & \textual \textasciitilde, T & -- & -- \\
         Limitations & \visual +, V & \textual +, T & \visual +, V & \textual +, T & \visual \textasciitilde, V & \textual \textasciitilde, T & \visual \textasciitilde, V & \textual \textasciitilde, T & -- & -- \\
         \bottomrule
    \end{tabular}
    }
    \label{tab:experimental_conditions}
    \vspace*{-0.5\baselineskip}
\end{table}

\paragraph*{Experimental Conditions (EC)}

The subset of experimental conditions selected for this user study is summarized in~Table~\ref{tab:experimental_conditions}. The conditions vary along three main dimensions: (1) response quality, (2) explanation quality, and (3) presentation of explanation. EC1 and EC2 represent a perfect system response with accurate explanations. 
More specifically, the explanations cover the source supporting the response, as well as the system's confidence score; the limitation explanation is not included because the response has no inaccuracies in this case. EC3 and EC4 correspond to imperfect responses that may contain some factual errors or be biased towards one specific point of view, but are accompanied by accurate explanations related to source, confidence, and limitations. EC5 and EC6 represent the perfect response accompanied by noisy explanations. EC7 and EC8 correspond to imperfect responses that contain flaws and noisy explanations. The last group of conditions, EC9 and EC10, represents the response (either perfect or imperfect) without explanations.

\subsection{Pilot Study}
\label{sec:pilot}

We ran a pilot study (MTurk; $N$=15; 3 HITs; US\$3 per HIT, proportional to US minimum wage), where HITs corresponded to three experimental conditions selected from the 10 described in Table~\ref{tab:experimental_conditions}: EC3, EC4, and EC7. 
The selected conditions encompass border cases, featuring variations in both the presentation mode of explanations (EC3 vs. EC4) and the quality explanations (EC3 vs. EC7), and deliberately involve imperfect responses to simulate the most natural scenarios. 
In their overall feedback, crowd workers primarily expressed concerns about the length of the task and the payment which was accordingly increased in the large-scale data collection.

We performed a power analysis
by employing one-way ANOVA with the experimental condition as an independent variable and user-reported response usefulness as a dependent variable
\citep{Sakai:2018:Book}. 
The results indicate that 16 workers are required to observe a statistically significant effect of explanation quality on the perceived usefulness of system responses, whereas 56 workers are required for a statistically significant effect of the explanation presentation mode. Considering four additional pairs of experimental conditions with varying presentation modes, we expect that gathering data from 14 unique workers per HIT (56 from power analysis divided by 4 pairs of conditions) is adequate to observe a statistically significant effect of presentation mode across all ten experimental conditions. 
Based on this analysis, we decided to recruit 16 unique workers per HIT in our main study.

\subsection{Main Study}
\label{sec:on_scale}

Crowd workers with a greater than 97\% approval rate, over 5,000 approved HITs, and located in the US were qualified to participate in the study. Workers were paid US\$4 for successful HIT completion. Workers who failed 4 out of 10 attentiveness checks or more were rejected. Altogether we collected 273 submissions, out of which 113 were discarded due to failed attentiveness checks. Accepted tasks were submitted by 160 unique workers (16 per HIT), with the following user-reported demographics: 95 male, 60 female, 5 in ``other'' category (none in ``prefer not to say''); age breakdown: 18--30 (39), 31--45 (76), 46--60 (41), 60+ (4); highest degree: Ph.D. or higher (3), master's (34), bachelor's (111), high school (12).
\section{Results}
\label{sec:results}

To answer our research questions, we first analyze the sensitivity of our experiment. Tables~\ref{tab:one-way-anova} and~\ref{tab:two-way-anova} show the results of the one- and two-way ANOVA tests for statistical significance on user-reported dimensions, using a significance level of $\alpha=0.05$.
Whenever applicable, the effect size of a given factor is classified based on the formula for the unbiased estimator and scales used by~\citet{Culpepper:2022:ACM}. Given the large number of factors defining each experimental condition, we treat response quality, quality of explanations, and their presentation mode as three separate independent variables to simplify the interpretation of the results. Each user-reported response dimension score and user rating for explanation is treated as a dependent variable. 
The analysis performed to answer RQ1 (Section~\ref{sec:results_quality}) and RQ2 (Section~\ref{sec:results_presentation}) is based only on the results with the statistically significant effects discussed in~Section~\ref{sec:results_sensitivity}.
\begin{table*}[tp]\centering
    \captionsetup{skip=2pt}
    \caption{Results of one-way ANOVA. Self-reported response dimensions (dependent variables) are in  columns, independent variables in rows. Boldface indicate statistically significant effects ($p<0.05$). Effect size: L=Large, M=Medium, S=Small.}
    \label{tab:one-way-anova}
    \sisetup{
    detect-weight=true,
    table-format=2.3,
    table-auto-round=false}
    \adjustbox{max width=1 \textwidth}{
    \begin{tabular}{lSSSSSSSSSSSSSSS@{\qquad}}
    \toprule
    & {\multirow{2}{*}{\bf Usefulness}} & \multicolumn{11}{c}{\bf Other Dimensions} & \multicolumn{3}{c}{\textbf{Explanation}}\\ 
  \cmidrule{3-13}\cmidrule(l){14-16}
     & & \textbf{Rel.} & \textbf{Correct.} & \textbf{Compl.} & \textbf{Comprehen.} & \textbf{Conciseness} & \textbf{Serendipity} & \textbf{Coherence} & \textbf{Factuality} & \textbf{Fairness} & \textbf{Read.}  &\textbf{Sat.} & \textbf{Source} & \textbf{Conf.} & \textbf{Limitation} \\
    \midrule
    & \multicolumn{15}{l}{\it All conditions (EC1--EC10)} \\
    \midrule
        Response Quality & 0.156 { (S)} &
        0.176 { (S)} & \bfseries0.003 { (S)} & 0.745 { (--)} & 0.846 { (--)} & 0.374 { (S)} & 0.093 { (S)} & 0.217 { (S)} & 0.265 { (S)} & 0.924 { (--)} & 0.881 { (--)} & 0.638 { (S)} & 0.697 { (--)} & 0.456 { (S)} & 0.445 { (S)} \\
        Explanation Quality & \bfseries0.0 { (S)} &
        \bfseries0.0 { (S)} & 0.508 { (S)} & \bfseries0.003 { (S)} & \bfseries0.0 { (S)} & \bfseries0.001 { (S)} & 0.09 { (S)} & \bfseries0.002 { (S)} & 0.713 { (--)} & \bfseries0.0 { (S)} & \bfseries0.032 { (S)} & \bfseries0.0 { (S)} & \bfseries0.0 { (S)} & \bfseries0.0 { (S)} & 0.173 { (S)} \\
        Presentation Mode & \bfseries0.019 { (S)} &
        \bfseries0.0 { (S)} & 0.234 { (S)} & 0.347 { (S)} & 0.658 { (--)} & \bfseries0.001 { (S)} & 0.149 { (S)} & 0.09 { (S)} & 0.842 { (--)} & \bfseries0.001 { (S)} & 0.651 { (--)} & \bfseries0.0 { (S)} & \bfseries0.0 { (S)} & \bfseries0.0 { (S)} & \bfseries0.0 { (S)} \\
        \midrule
        Query  & 0.341 { (S)} & 
        0.911 { (--)} & 0.939 { (--)} & 0.84 { (--)} & 0.733 { (--)} & 0.449 { (S)} & 0.66 { (--)} & 0.543 { (--)} & 0.724 { (--)} & 0.098 { (S)} & 0.125 { (S)} & 0.254 { (S)} & 1.0 { (--)} & 1.0 { (--)} & 1.0 { (--)} \\
        \midrule
        Topic Familiarity & \bfseries0.017 { (S)} &
        \bfseries0.0 { (S)} & 0.285 { (S)} & \bfseries0.0 { (S)} & \bfseries0.0 { (S)} & \bfseries0.0 { (S)} & \bfseries0.0 { (M)} & \bfseries0.0 { (S)} & \bfseries0.0 { (S)} & \bfseries0.0 { (S)} & \bfseries0.002 { (S)} & \bfseries0.0 { (S)} & \bfseries0.0 { (M)} & \bfseries0.0 { (S)} & \bfseries0.0 { (S)} \\
        Interest In Topic & \bfseries0.0 { (S)} &
        \bfseries0.007 { (S)} & \bfseries0.0 { (S)} & \bfseries0.0 { (S)} & \bfseries0.0 { (S)} & 0.053 { (S)} & \bfseries0.0 { (M)} & 0.115 { (S)} & \bfseries0.0 { (S)} & \bfseries0.0 { (S)} & \bfseries0.0 { (S)} 
        & \bfseries0.0 { (S)} & \bfseries0.0 { (M)} & \bfseries0.0 { (S)} & \bfseries0.0 { (S)} \\        
        Similar Search Prob. & \bfseries0.0 { (S)} & 
        \bfseries0.0 { (S)} & \bfseries0.001 { (S)} & \bfseries0.0 { (M)} & \bfseries0.0 { (S)} & \bfseries0.0 { (S)} & \bfseries0.0 { (M)} & 
        \bfseries0.002 { (S)} & \bfseries0.0 { (S)} & \bfseries0.0 { (S)} & \bfseries0.0 { (S)} & \bfseries0.0 { (S)} & \bfseries0.0 { (M)} & \bfseries0.0 { (S)} & \bfseries0.0 { (S)} \\
     \midrule
        Conv. Agent Familiarity  & 0.079 { (S)} &
        \bfseries0.0 { (S)} & 0.077 { (S)} & \bfseries0.001 { (S)} & \bfseries0.0 { (S)} & 0.093 { (S)} & \bfseries0.0 { (S)} & \bfseries0.003 { (S)} & \bfseries0.0 { (S)} & 0.079 { (S)} & \bfseries0.005 { (S)} & \bfseries0.004 { (S)} & \bfseries0.0 { (S)} & \bfseries0.0 { (S)} & \bfseries0.0 { (S)} \\
        Search with Agent Freq. & \bfseries0.0 { (S)} &
        \bfseries0.002 { (S)} & 0.351 { (S)} & \bfseries0.0 { (S)} & \bfseries0.0 { (S)} & \bfseries0.0 { (S)} & \bfseries0.0 { (M)} & \bfseries0.0 { (S)} & 0.533 { (--)} & 0.426 { (S)} & \bfseries0.0 { (S)} & \bfseries0.0 { (S)} & \bfseries0.0 { (M)} & \bfseries0.0 { (S)} & \bfseries0.0 { (M)} \\
    \midrule
    
    & \multicolumn{15}{l}{\it Only conditions with explanations (EC1--EC8)} \\
    \midrule
        Explanation Quality & \bfseries0.0 { (S)} & \bfseries0.006 { (S)} & 0.256 { (S)} & \bfseries0.002 { (S)} & \bfseries0.0 { (S)} & 0.122 { (S)} & 0.319 { (S)} & \bfseries0.003 { (S)} & 0.504 { (S)} & \bfseries0.0 { (S)} & \bfseries0.014 { (S)} & \bfseries0.007 { (S)} & 0.097 { (S)} & \bfseries0.0 { (S)} & 0.088 { (S)} \\
        Presentation Mode & 0.872 { (--)} & 0.686 { (--)} & 0.096 { (S)} & 0.895 { (--)} & 0.38 { (S)} & 0.399 { (S)} & 0.86 { (--)} & 0.377 { (S)} & 0.739 { (--)} & 0.78 { (--)} & 0.771 { (--)} & 0.071 { (S)} & \bfseries0.0 { (S)} & 0.653 { (--)} & \bfseries0.0 { (S)} \\
    \bottomrule
\end{tabular}
}
\vspace{-2mm}
\end{table*}
\begin{table}[tp]\centering
    \captionsetup{skip=2pt}
    \caption{Results of two-way ANOVA. Boldface indicates statistically significant effects ($p<0.05$). Effect size: S=Small.}
    \label{tab:two-way-anova}
      \sisetup{
    detect-weight=true,
    table-format=2.3,
    table-auto-round=false}
   \adjustbox{max width=0.48\textwidth}{
    \begin{tabular}{lSSSSS@{\qquad}}
    \toprule
& {\multirow{2}{*}{\textbf{Usefulness}}} & {\multirow{2}{*}{\textbf{Satisfaction}}} & \multicolumn{3}{c}{\bf Explanation} \\
& & & \textbf{Source} & \textbf{Confidence} & \textbf{Limitation} \\
     
    \midrule
    & \multicolumn{5}{l}{{\it Interactions with Query}} \\
    \midrule
        Response Quality & 
        0.069 { (S)} & 0.296 { (S)} &
        1.0 { (--)} & 1.0 { (--)} & 1.0 { (--)} \\
        Explanation Quality & 
        0.767 { (--)} & 0.993 { (--)} & 1.0 { (--)} & 1.0 { (--)} & 1.0 { (--)} \\
        Presentation Mode & 0.94 { (--)} & 0.981 { (--)} & 1.0 { (--)} & 1.0 { (--)} & 1.0 { (--)} \\
        \midrule
        Conv. Agent Familiarity & 0.995 { (--)} & 0.887 { (--)} & 1.0 { (--)} & 1.0 { (--)} & 1.0 { (--)} \\
        Search with Agent Freq. & 0.632 { (--)} & 0.215 { (S)} & 1.0 { (--)} & 1.0 { (--)} & 1.0 { (--)} \\
        \midrule
        Topic Familiarity & 0.697 { (--)} & 0.489 { (S)} & \bfseries0.002 { (S)} & 0.71 { (--)} & \bfseries0.001 { (S)} \\
        Interest in Topic & 0.087 { (S)} & 0.542 { (--)} & 0.063 { (S)} & 0.698 { (--)} & 0.234 { (S)} \\
        Similar Search Prob. & \bfseries0.014 { (S)} & \bfseries 0.019 { (S)} & 0.449 { (S)} & 0.922 { (--)} & 0.082 { (S)} \\
    \midrule
    & \multicolumn{5}{l}{\it Interactions with Topic Familiarity} \\
    \midrule
        Response Quality & 0.848 { (--)} & 0.42 { (S)} & 0.24 { (S)} & \bfseries0.005 { (S)} & \bfseries0.0 { (S)} \\
        Explanation Quality & 0.155 { (S)} & 0.671 { (--)} & \bfseries0.0 { (S)} & \bfseries0.0 { (S)} & \bfseries0.0 { (S)} \\
        Presentation Mode & 0.663 { (--)} & 0.752 { (--)} & \bfseries0.0 { (S)} & \bfseries0.0 { (S)} & \bfseries0.0 { (S)} \\
    \bottomrule
\end{tabular}
}
\vspace{-4mm}
\end{table}

\subsection{User's Perception of Response and Explanations}
\label{sec:results_sensitivity}

\paragraph*{Response Quality.} Table~\ref{tab:one-way-anova} shows that \emph{response quality} has a statistically significant effect only on user-reported correctness of the response. Completeness, factuality, and fairness are not influenced by the quality of the response, even though some responses contained manually injected errors related to these dimensions (e.g., bias towards one specific point of view, factual errors, or covering only one aspect of the topic). This insensitivity of user-reported response dimensions to the quality of provided information may suggest that users are not able to identify some of the problems with the response without expert knowledge about the topic. 

\paragraph*{Explanations.}
Our experiments include two experimental conditions where explanations are not provided (i.e., EC9--EC10). In order to understand the impact of quality and presentation mode of explanations, we conducted an additional analysis on the data from HITs representing only EC1--EC8 (reported in the bottom part of  Table~\ref{tab:one-way-anova}) and we focused our analysis on these results.
In terms of \emph{explanation quality}, we observe that introducing noise in explanations has a statistically significant effect on almost all user-reported response dimensions, suggesting that noisy explanations have a strong impact on user experience in general. However, the quality of explanations does not impact user assessment of correctness and factuality, dimensions related to factual errors in the response. It means that users seem to assess the factual correctness of the response independently of the quality of the explanations provided by the system. 
In terms of \emph{presentation mode}, we observe a statistically significant effect only for the presence/absence of explanations on the usefulness of the response---a statistically significant effect is observed in the top part but not in the bottom part of the table. Similarly, the user-reported conciseness, fairness, and relevance of the response are impacted only by the presence/absence of explanations. This implies insensitivity of the response dimensions to the way explanations are presented. 

\paragraph*{User Ratings for Source, Confidence, and Limitations.}
The impact of noise in the source is solely tied to the presence or absence of the source---no statistically significant effect is observed for EC1--EC8. However, the presentation mode of the source affects user ratings for the explanations independently of its presence or absence (statistical significance persists when excluding EC9 and EC10), even though the source is presented in the same format in both presentation modes. This may be due to the wording of the question about source explanation in the questionnaire---it does not explicitly mention sources, and therefore is open for other interpretation, especially when sources are not provided. 
In the case of noise in confidence explanation, it significantly affects user ratings. However, concerning presentation mode, we can only discern the effect of its presence or absence, not the specific mode of presentation.
Regarding limitations, there is no statistically significant effect of noise in the corresponding explanation, but there is of presentation mode. User ratings for explanations related to limitations are influenced by the presentation mode, not the mere presence of this explanation.
This implies that, in general, the impact of noise on explanations is only related to the confidence and the impact of the presentation mode only to the limitations. The effect of quality and presentation mode on other explanations---based on the user ratings---was not significant in this user study. 

\paragraph*{Query.} We do not observe any statistically significant effects of the query on the user-reported response dimensions. This suggests that the results are topic-independent and generalizable. The proposed user study design mitigates the impact of the query on the results. Additionally, the interaction between response quality, quality of explanations, or presentation mode and the query does not have a statistically significant effect on user-reported scores for response satisfaction, usefulness, and explanations (see~Table~\ref{tab:two-way-anova}).

\paragraph*{Topic Familiarity.} Workers report that they are rather familiar with the query topics, which indicates that the process of query selection was successful. Following our hypothesis, users' background knowledge about the topic affects how they assess the response. It is visible on the effects reported for almost all response dimensions (see~Table~\ref{tab:one-way-anova}).
Similar effects are observed for the user's interest in the topic and the likelihood of the user searching for a similar query. Additionally, we observe a statistically significant effect of all these three indicators of user involvement in the topic on the user ratings for explanations. It implies that these factors that we cannot control and are completely user-dependent directly impact the assessment of the responses we examined in this user study. 

In terms of the results of two-way ANOVA (see~Table~\ref{tab:two-way-anova}), we observe a statistically significant effect of the interaction between the user's familiarity with the topic and the response quality on the user ratings for explanations related to limitation and the system's confidence. It confirms the intuitive relationship between the user's background knowledge and the quality of the response on their ability to correctly assess the explanations provided by the system and deem it useful or not. We do not observe a statistically significant effect of interaction between response quality and familiarity with the topic on the usefulness of the response or user satisfaction in general.
The interaction between the noise in the explanations and the familiarity with the topic has a statistically significant effect on the user ratings for all three explanations. It can follow from the fact that a user unfamiliar with the topic needs high-quality explanations from the system to be able to verify and use the provided response.
The user ratings for explanations are influenced by the interaction between the way explanations are presented and the user's familiarity with the topic. This suggests that depending on the user's background knowledge, the preferred way of receiving explanations from the system may differ. 

\paragraph*{Familiarity with Conversational Agents.}
We observe a statistically significant effect of the user-reported frequency of interacting with conversational assistants (in general and for search specifically) on some of the user-reported response dimensions (see~Table~\ref{tab:one-way-anova}). Interestingly, we observe a medium-size effect of the frequency of using the conversational search on the user ratings for explanations related to source and limitations. Additionally, we observe that higher values for familiarity with conversational agents are associated with explanations without noise and visual presentation mode. It indicates that the user's familiarity with the system impacts their assessment of its additional components. 

\subsection{Effect of the Explanation Quality (RQ1)}
\label{sec:results_quality}

\paragraph*{Effect on the User-reported Response Dimensions}

\begin{figure}[tp]
    \centering
    \vspace*{-0.25\baselineskip}
    \hspace{-0.2cm}
    \includegraphics[width=0.48\textwidth]
    {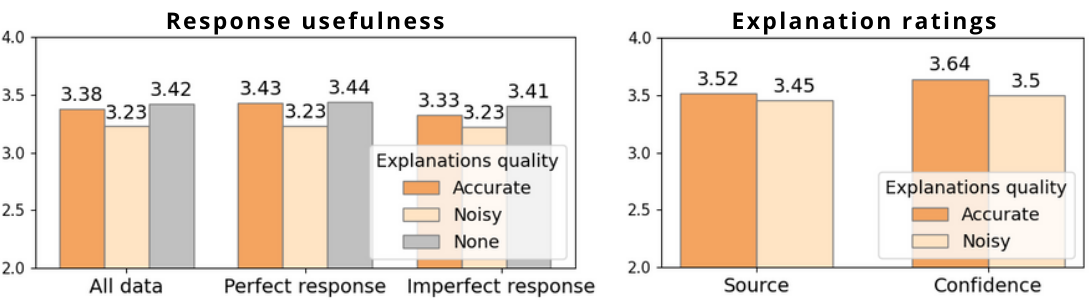}%
    \\
    \hspace{-0.2cm}
    \includegraphics[width=0.48\textwidth]{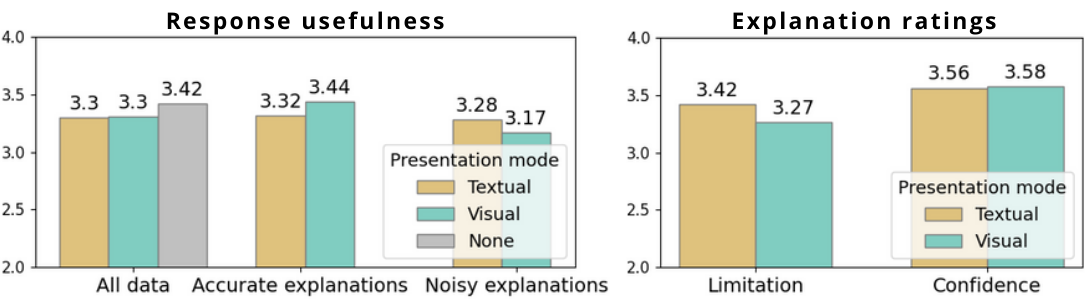}
    \captionsetup{aboveskip=8pt}
    \caption{Mean scores for response usefulness and explanation ratings for different quality of the explanations (top) and presentation mode (bottom). All differences between the ratings within a given plot are statistically significant.
    }
    \label{fig:results:barplots}
\end{figure}

The top-left plot in~Figure~\ref{fig:results:barplots} shows that user-reported values for the usefulness of the responses are concentrated around higher values (3 and 4). However, noise in the explanations results in slightly lower usefulness scores. It indicates that the high-quality source, system confidence score, and information about the response limitations make the response more useful from the user's perspective.
Minor differences in usefulness scores between perfect and imperfect responses (second and third set of bars in the plot) suggest that when explanations are not provided (``None'' variant), users are less likely to object to the usefulness of imperfect responses. 
In general, the explanations are meant to increase the reliability and transparency of the system. However, they require additional time and effort from the user and the cost of ``processing'' explanations may be higher than the actual gain. This situation is visible in the second and third set of bars in the top-left plot in~Figure~\ref{fig:results:barplots} where the highest usefulness is reported for the responses that do not contain any explanations (``None'' variant), independent of their quality. It suggests that the explanations either pollute the response or make the user more critical about it, resulting in reduced usefulness. 

\paragraph*{Effect on user ratings of explanations}
Looking at the means of user ratings for source and confidence explanations (top-right plot in~Figure~\ref{fig:results:barplots}), ratings are again skewed towards higher values, and scores for accurate explanations are slightly higher than for noisy explanations, especially for confidence. This suggests that users perceive noisy explanations as less useful in understanding system confidence and attributed sources---we do not observe statistically significant differences for the explanations related to limitations. 

\subsection{Effect of the 
Presentation Mode (RQ2)}
\label{sec:results_presentation}

\paragraph*{Effect on the user-reported response dimensions}
On average, we do not observe differences in the usefulness scores between textual and visual modes, but usefulness scores are significantly higher when no explanations are provided (bottom-left plot in~Figure~\ref{fig:results:barplots}). This is aligned with the one-way ANOVA results and suggests that the main issue is not the question of presentation mode but rather whether the explanations are necessary, hinting at the underlying trade-off between effort and gain.
Nevertheless, we observe some differences in the user ratings for explanations when looking at responses accompanied by explanations with different quality. Namely, visual explanations result in higher usefulness scores for responses with accurate explanations, while in case of noisy explanations workers find the textual format more useful. 

\paragraph*{Effect on user ratings of explanations}
Looking at the means of user ratings for explanations with respect to different presentation modes (bottom-right plot in~Figure~\ref{fig:results:barplots}), the preferred presentation mode depends on the explained aspect of the response. (Note that user ratings for explanations related to the source are not informative in this case, as the source is always presented in the same way.) Namely, we observe slightly higher ratings for the textual presentation of limitations. In the case of confidence, the difference between presentation modes is very small with a slight preference towards visual presentation, which aligns with the results of one-way ANOVA. This suggests that further research is needed to better understand how to optimally integrate different aspects in the layout of transparent CIS responses.

\subsection{Qualitative Analysis}

We manually investigate the feedback given by crowd workers regarding their ratings for the source, confidence, and limitations explanations, seeking insights and suggestions to enhance their content and presentation.\footnote{Comments provided by workers can be found in the online repository.} 
Many workers (18/160) pointed out that explanations related to limitations and confidence significantly enhanced their understanding of the constraints of both the system and the responses. The mention of encouragement towards information verification and critical thinking was consistent across various qualities (comments from EC1--EC8 HITs), and positive comments were also shared for noisy explanations (EC5--EC8). It suggests that workers may face challenges in identifying inaccuracies in the explanations. 
For instance, even though the provided sources did not align with the information in the response, none of the users mentioned these mismatches in their comments.
Nevertheless, several crowd workers (4/160) emphasized the potential insufficiency of responses restricted to three sentences and a single source in certain situations.
A few (3/160) crowd workers expressed uncertainty in interpreting explanations related to limitations and confidence scores, underscoring the need for additional explanations or tutorials describing the system interface before usage. For instance, some workers attempted to interpret the meaning of the confidence score on their own describing it as a ``transparency measure to indicate the system’s level of certainty regarding the accuracy or relevance of the information shared'' or a ``model's estimate of the accuracy and reliability of its responses.'' 
In terms of the presentation mode, one worker suggested that representing confidence score using percentages would be more precise and helpful than a ``wifi connection symbol.'' This suggests that users might prefer a different display element, e.g., a fuel gauge~\cite{Shani:2013:Ja}, and perhaps also a finer confidence scale (which would require a more precise estimation of confidence).
In HITs with no explanations (EC9 and EC10), workers highlighted their lack of awareness regarding response limitations and confidence. Some workers attempted to gauge system confidence by searching for implicit confidence signals like ``I think'' or ``I believe'' in the responses~\citep{Radensky:2023:FAccT}. 
Overall, workers consistently emphasized that explanations enhance their understanding and encourage information verification and critical thinking. However, the comments reflect that workers are unlikely to identify flaws in the provided explanations.
\section{Discussion}
\label{sec:discussion}

Our results show that high-quality explanations related to the source, system confidence, and response limitations increase the user-perceived usefulness of the response and user ratings for explanations. Additionally, noise in the explanations of the response provided by the system has a significant impact on user experience in general (almost all response dimensions are affected). These results align with previous research in AI-assisted decision-making claiming that confidence scores can help calibrate people’s trust in the system model, but they are not sufficient to increase the success rate of interactions~\citep {Zhang:2020:FAT}. 
In our study, we observe a significant effect of familiarity with the topic on response assessment, indicating the need for the user's background knowledge to complement the system’s errors~\citep{Zhang:2020:FAT}.
In terms of user's sensitivity to inaccuracies in the responses and explanations, we show that users are not able to detect factual errors or biases in the provided information. The qualitative analysis shows that workers do not point out these inaccuracies explicitly.
Similarly, they cannot identify flaws in the explanations related to response limitations. This aligns with previous research, demonstrating that explanations might cause users to follow the system’s advice more often, even when it is wrong~\citep{vanderWaa:2021:Artificial}.
Our study is not conclusive about the preferred way of presenting explanations to the user. We find that limitations tend to receive higher user ratings when presented in a textual form, whereas, for confidence, we observe the opposite trend, which complies with the findings reported in the field of recommender systems~\citep{Shani:2013:Ja}. Additionally, limited mentions of the presentation mode in the free-text feedback obtained from crowd workers may imply that the format of explanations is not a crucial factor in this setting.

Insights from this study about communicating explanations to facilitate users' assessment of the provided information need to be put in a broader context of system explainability and the associated effort/gain trade-off~\citep{Cheng:2019:CHI}. While these explanations complement the system response with components that enable users to assess responses more objectively, they demand more time and effort than merely reading the provided response.
Optimizing user gain is a complex task influenced by various factors. Firstly, the relationship between the user's gain and the effort associated with the amount of additional information is not linear; while more explanations generally increase gain, there is a tolerance threshold. Exceeding that threshold may overwhelm users, causing a drop in gain.
Secondly, the overall quality of the system's response and explanations significantly impacts gain. This is evidenced by our findings: users struggle to detect flaws in provided responses when explanations contain noise or errors, and providing no explanations is more useful than providing noisy ones. Thirdly, the relevance of explanations depends on the topic's complexity and user familiarity, with more complex topics benefiting from adjusted and detailed information. 
Additionally, the optimal effort-gain trade-off is likely to be user-dependent, requiring personalized adjustments in the amount, level of detail, and presentation of the information, which is evidenced by various preferences for the confidence display we observed in the feedback. 
To our knowledge, investigating the adaptation of responses based on user preferences, previous interactions with CIS systems, and topic complexity has yet to be explored.
\section{Conclusion}
\label{sec:concl}

Response transparency has not received significant attention in a CIS setting. Our user study addresses this gap by examining various ways of explaining the source of the information provided by the system, the system's confidence in the response, and its limitations. We explore the effect of noise and different presentation modes of these explanations on users' assessments of responses and explanations. 
Results reveal lower user-reported scores when explanations contain noise, although these scores seem insensitive to the quality of the response. 
In terms of presentation mode, we do not observe significant differences between visual and textual explanations---suggesting that the format of explanations may not be a critical factor in this setting---but users presented with no explanations found the responses more useful.
To our knowledge, this study is the first to examine response transparency in CIS, highlighting the need for further research to enhance transparency in CIS responses. In particular, some of the limitations of our experimental design can be addressed by studying the impact of response specificity and interactivity on user experience over time, analyzing user's assessment when provided with a broader context or previous interactions%
, and evaluating in more detail the relation between user's background knowledge and the usefulness of explanations.

\section*{Acknowledgments}

This research was supported by the Norwegian Research Center for AI Innovation, NorwAI (Research Council of Norway, project number 309834), and by the \grantsponsor{ARC}{Australian Research Council}{https://www.arc.gov.au/} (\grantnum{ARC}{DE200100064}, \grantnum{ARC}{CE200100005}).

\balance
\bibliographystyle{ACM-Reference-Format}
\bibliography{sigir2024-res_gen.bib}

\end{document}